\documentclass[]{article}
\usepackage{amsmath, amssymb, amsthm,bbm}
\usepackage{algorithm, algorithmic}
\usepackage{geometry}
\RequirePackage[numbers]{natbib}

\usepackage{multirow}
\usepackage{booktabs}
\usepackage{orcidlink}

\theoremstyle{plain}

\title{Robust reduced rank regression under heavy-tailed noise and missing data via non-convex penalization}
\author{The Tien Mai\orcidlink{0000-0002-3514-9636}}

\date{
\small
Norwegian Institute of Public Health, Oslo, 0456, Norway
\\
email: the.tien.mai@fhi.no
}

\begin{document}

\maketitle

\begin{abstract}
    Reduced rank regression (RRR) is a fundamental tool for modeling multiple responses through low-dimensional latent structures, offering both interpretability and strong predictive performance in high-dimensional settings. 
    Classical RRR methods, however, typically rely on squared loss and Gaussian noise assumptions, rendering them sensitive to heavy-tailed errors, outliers, and data contamination. 
    Moreover, the presence of missing data—common in modern applications—further complicates reliable low-rank estimation. In this paper, we propose a robust reduced rank regression framework that simultaneously addresses heavy-tailed noise, outliers, and missing data. Our approach combines a robust Huber loss with nonconvex spectral regularization, specifically the minimax concave penalty (MCP) and smoothly clipped absolute deviation (SCAD).
    Unlike convex nuclear-norm regularization, the proposed nonconvex penalties alleviate excessive shrinkage and enable more accurate recovery of the underlying low-rank structure.
    The method also accommodates missing data in the response matrix without requiring imputation. We develop an efficient proximal gradient algorithm based on alternating updates and tailored spectral thresholding. Extensive simulation studies demonstrate that the proposed methods substantially outperform nuclear-norm–based and non-robust alternatives under heavy-tailed noise and contamination. 
An application to cancer cell line data set further illustrates the practical advantages of the proposed robust RRR framework.

Our method is implemented in the \texttt{R} package \texttt{rrpackrobust} available at \url{https://github.com/tienmt/rrpackrobust}.
\end{abstract}


\section{Introduction}

Characterizing the structural relationships between multiple outcomes and a common set of predictors is a fundamental objective across a wide range of scientific and applied disciplines, including genomics, finance, signal processing, and the social sciences \citep{izenman2008modern,cook2018introduction,giraud2021introduction}. This long-standing methodological challenge has motivated extensive research on multivariate modeling, with particular emphasis on uncovering latent structure and improving predictive performance in high-dimensional systems. Among the resulting methodologies, reduced rank regression (RRR) has emerged as a central framework for jointly modeling multiple responses through a low-dimensional representation of the coefficient matrix \citep{anderson1951estimating,izenman1975reduced,reinsel2023multivariate}. By imposing a rank constraint, RRR achieves effective dimension reduction, enhances interpretability via latent factors, and enables stable estimation in settings with large numbers of predictors or responses \citep{bunea2011optimal,bunea2012joint}. A wide range of estimation and inference strategies for RRR have been proposed \citep{geweke1996bayesian,chen2013reduced,she2017robust,bing2019adaptive,goh2017bayesian,mai2023reduced,mai2023bilinear,mai2025concentration,mai2025properties,chakraborty2020bayesian}.

Classical RRR formulations are typically built upon squared loss, implicitly assuming Gaussian noise. However, this assumption is often violated in real-world applications, where responses may be contaminated by outliers or arise from inherently heavy-tailed distributions such as Student’s (t) or impulsive mixture models. In such settings, ordinary least squares–based estimators can exhibit severe degradation in both estimation accuracy and predictive performance, as even a small number of anomalous observations can distort the estimated low-rank structure \citep{huber1996robust}. This vulnerability is particularly pronounced in low-rank matrix estimation, where outliers may inflate singular values or corrupt latent factors, rendering standard nuclear-norm–regularized RRR unstable and biased.

Missing data further complicate the problem. In many applications, the multiple response matrix is only partially observed due to measurement failures, censoring, or data integration issues. Although several methods have been developed to address incomplete responses in RRR \citep{luo2018leveraging,mai2023bilinear,mai2023reduced}, these approaches rely on squared loss and are not designed to handle heavy-tailed noise or outliers. To date, there is a lack of robust reduced rank regression methods that simultaneously accommodate missing responses and non-Gaussian contamination.

In this work, we address these challenges by proposing a robust reduced rank regression framework that combines the Huber loss with nonconvex spectral regularization. Specifically, we employ MCP and SCAD penalties applied directly to the singular values of the coefficient matrix, enabling adaptive low-rank estimation with reduced bias relative to convex nuclear-norm penalization. While existing robust RRR methods often rely on convex relaxations \citep{tan2024adaptive,ma2024d4r} or explicit mean-shift modeling of outliers \citep{she2017robust}, such approaches may overshrink large singular values or introduce additional modeling complexity, particularly in the presence of missing data. In contrast, nonconvex spectral penalties preserve large singular values while shrinking small ones toward zero, yielding improved rank recovery and estimation accuracy in high-dimensional regimes.

To enhance robustness against heavy-tailed errors and outliers, we integrate the Huber loss into the RRR framework. This choice provides resistance to contamination without introducing auxiliary nuisance parameters, while naturally extending to settings with incomplete response matrices by evaluating the loss only on observed entries. The resulting optimization problem is nonconvex due to the spectral penalty; nevertheless, we develop an efficient proximal gradient algorithm that alternates between gradient updates for the smooth Huber loss and closed-form proximal mappings for the MCP and SCAD penalties. Each iteration reduces to singular value thresholding with firm-shrinkage rules, enabling scalable computation via truncated or randomized SVDs.

We illustrate the practical effectiveness of the proposed method through extensive simulation studies and real-data applications. In simulations, our methods consistently outperform nuclear-norm–based and non-robust competitors in terms of estimation accuracy, rank recovery, and prediction error, particularly under heavy-tailed noise and increasing levels of missingness. Notably, compared with the mRRR method of \cite{luo2018leveraging}, which is designed for incomplete responses but relies on squared loss, our robust approach achieves substantial gains when the data are contaminated. Real-data analyses to NCI-60 cancer cell line data set further demonstrate the practical advantages of our framework, confirming its ability to deliver stable and interpretable low-rank estimates in challenging, noisy, and incomplete environments.

The remainder of the paper is organized as follows. Section~\ref{sc_method} introduces the problem formulation and the proposed methodology. The algorithmic development is detailed in Section~\ref{sc_algorithm}. Simulation studies are presented in Section~\ref{sc_simulations}, and a real data application is discussed in Section~\ref{sc_application}. Section~\ref{sc_concluson} concludes with a discussion and final remarks. Our method is implemented in the \texttt{R} package \texttt{rrpackrobust} available at \url{https://github.com/tienmt/rrpackrobust}.

\section{Problem and Method}
\label{sc_method}
\subsection{Model Formulation}

We consider a multiple-response regression setting with response matrix 
$\mathbf{Y} \in \mathbb{R}^{n \times q}$ and predictor matrix 
$\mathbf{X} \in \mathbb{R}^{n \times p}$. 
The goal is to estimate a low-rank coefficient matrix 
$\mathbf{B} \in \mathbb{R}^{p \times q}$ that captures the shared latent structure between predictors and responses while remaining robust to outliers or heavy-tailed errors.

We model
\begin{equation}
\label{eq_main_model}
    \mathbf{Y} = \mathbf{X}\mathbf{B} + \mathbf{E},
\end{equation}
where $\mathbf{E}$ denotes a  noise matrix whose entries are zero-mean and may not be sub-Gaussian. 
To achieve robustness, we replace the conventional least-squares loss with the 
\emph{Huber loss} \citep{huber1964robust}, defined element-wise by
\begin{equation}
    \rho_\tau(u) = 
\begin{cases}
\frac{1}{2}u^2, & \text{if } |u| \le \tau, \\
\tau|u| - \frac{1}{2}\tau^2, & \text{if } |u| > \tau,
\end{cases}
\end{equation}
where $\tau>0$ is a user-specified threshold controlling the transition between quadratic (for small residuals) and linear (for large residuals) regimes.

The corresponding optimization problem is
\begin{equation}
\label{eq:objective}
\widehat{\mathbf{B}} 
= 
\arg\min_{\mathbf{B} }
\Bigg\{
\frac{1}{n}\sum_{i=1}^{n} \sum_{j=1}^{q}
\rho_\tau\big( \mathbf{Y}_{ij} - \mathbf{X}_i^\top \mathbf{B}_j \big)
+ R_\eta(\mathbf{B})
\Bigg\},
\end{equation}
where $R_\eta(\mathbf{B})$ is a spectral regularizer promoting low-rankness.

\subsection{Nonconvex Spectral Regularization}

Let $\sigma_1(\mathbf{B}), \ldots, \sigma_{\min(p,q)}(\mathbf{B})$ denote the singular values of $\mathbf{B}$. 
We put
\[
R_\eta(\mathbf{B}) = \sum_{j=1}^{\min(p,q)} \rho_\eta(\sigma_j(\mathbf{B})),
\]
where $\rho_\eta$ is a nonconvex sparsity-inducing penalty such as
the smoothly clipped absolute deviation (SCAD; \citealp{fan2001variable}) or
the minimax concave penalty (MCP; \citealp{zhang2010nearly}).
Specifically, the SCAD penalty is defined as:
\begin{equation}
\label{eq_SCAD_penalty}
\rho_{\eta}^{\mathrm{SCAD}}(t) =
\begin{cases}
\lambda t, & 0 \le t \le \lambda, \\[4pt]
\dfrac{-t^2 + 2\eta\lambda t - \lambda^2}{2(\eta - 1)}, &
\lambda < t \le \eta\lambda, \\[8pt]
\dfrac{(\eta + 1)\lambda^2}{2}, & t > \eta\lambda,
\end{cases}
\; \text{with } \eta > 2.
\end{equation}
and the MCP penalty is defined as
\begin{equation}
\label{eq_MCP}
\rho_{\eta}^{\mathrm{MCP}}(t) =
\begin{cases}
\lambda t - \dfrac{t^2}{2\eta}, & 0 \le t \le \eta \lambda, \\[4pt]
\frac{1}{2}\eta\lambda^2, & t > \eta \lambda,
\end{cases}
\; \text{with } \eta > 1.
\end{equation}
where $\lambda>0$ controls regularization strength, and $ \eta>1$ determines the concavity of the penalty.
These penalties provide a nonconvex relaxation of the nuclear norm, yielding better bias–variance trade-offs and rank recovery properties.

\subsection{Handling missing data}
To accommodate missing entries in the multivariate response matrix, we adopt a masking strategy analogous to that used in matrix completion \citep{luo2018leveraging,mai2023reduced}.
Let \( \mathbf{Y} \in \mathbb{R}^{n \times q} \) denote the response matrix, 
and let \( \Omega \subset \{1,\ldots,n\} \times \{1,\ldots,q\} \) be the set of observed indices. The corresponding optimization problem from \eqref{eq:objective} becomes
\begin{equation}
\widehat{\mathbf{B}} 
= 
\arg\min_{\mathbf{B} }
\Bigg\{
\frac{1}{n}\sum_{(i,j)\in\Omega }
\rho_\tau\big( \mathbf{Y}_{ij} - \mathbf{X}_i^\top \mathbf{B}_j \big)
+ R_\eta(\mathbf{B})
\Bigg\}
.
\end{equation}

\section{Proximal Gradient Algorithm}
\label{sc_algorithm}
\subsection{Algorithm development}

Problem \eqref{eq:objective} is nonconvex due to the non-convex spectral penalty.  
We employ a proximal gradient algorithm \citep{parikh2014proximal}, which iteratively updates $\mathbf{B}$ via a gradient descent step on the smooth loss and a proximal mapping on the spectral regularizer.

Let
\[
L(\mathbf{B}) = \frac{1}{n}\sum_{i=1}^{n}
\rho_\tau(\mathbf{Y}_i - \mathbf{X}_i^\top \mathbf{B}),
\quad
\nabla L(\mathbf{B}) = -\frac{1}{n}\mathbf{X}^\top \psi_\tau(\mathbf{Y}-\mathbf{X}\mathbf{B}),
\]
where $ \; \psi_\tau(u) = \mathrm{sign}(u) \cdot \min(|u|, \tau)\; $ is the derivative of the Huber loss.

Given a step size $\alpha > 0$, the update rule is:
\begin{equation}
\label{eq:update}
\mathbf{B}^{(t+1)} 
= 
\mathrm{prox}_{\alpha R_\eta}\Big(
\mathbf{B}^{(t)} - \alpha \nabla L(\mathbf{B}^{(t)})
\Big),
\end{equation}
where $\mathrm{prox}_{\alpha R_\eta}(\cdot)$ denotes the proximal operator associated with $R_\eta$:
\[
\mathrm{prox}_{\alpha R_\eta}(\mathbf{M})
=
\arg\min_{\mathbf{B}}
\Big\{
\frac{1}{2}\|\mathbf{B}-\mathbf{M}\|_F^2 + \alpha R_\eta(\mathbf{B})
\Big\}.
\]
The step size $\alpha$ can be chosen as 
$\alpha = 1/L_X$, where 
$L_X = \|\mathbf{X}\|_2^2 / n$
is the Lipschitz constant of the gradient.
\\
Because $R_\eta$ is spectral (depends only on singular values), the proximal operator can be computed via singular value decomposition (SVD):
\[
\mathbf{M} = \mathbf{U} \, \mathrm{diag}(s_1, \ldots, s_r) \, \mathbf{V}^\top,
\quad
\mathrm{prox}_{\alpha R_\eta}(\mathbf{M})
= \mathbf{U} \, \mathrm{diag}\big(\mathrm{prox}_{\rho_\eta}^\text{scalar}(s_j)\big) \, \mathbf{V}^\top,
\]
where $\mathrm{prox}_{\rho_\eta}^\text{scalar}$ is the scalar proximal mapping under SCAD or MCP.

\subsection{Proximal operator of the nonconvex spectral penalty}

Consider the nonconvex spectral regularizer
\[
R_\eta(B) = \sum_{j=1}^{\min(p,q)} \rho_\eta(\sigma_j(B)),
\]
where $\sigma_j(B)$ denotes the $j$th singular value of the matrix $B$. The associated proximal operator is given by
\[
\mathrm{prox}_{\lambda R_\eta}(B)
= \arg\min_{A \in \mathbb{R}^{p \times q}}
\frac{1}{2}\|A - B\|_F^2 + \lambda R_\eta(A),
\]
which can be evaluated through singular value thresholding:
\[
\mathrm{prox}_{\lambda R_\eta}(B)
= U\,\mathrm{diag}\!\left(
\mathrm{prox}_{\rho_\eta}^{\text{scalar}}(\sigma_j(B); \lambda)
\right)V^\top,
\]
where $B = U \Sigma V^\top$ is the singular value decomposition (SVD) of $B$, and
\[
\mathrm{prox}_{\rho_\eta}^{\text{scalar}}(z; \lambda)
= \arg\min_{x \in \mathbb{R}} \frac{1}{2}(x - z)^2 + \lambda \rho_\eta(|x|)
\]
is the \emph{scalar proximal operator} associated with $\rho_\eta(\cdot)$.
This operator acts elementwise on the singular values and generalizes soft-thresholding to nonconvex shrinkage rules.

\paragraph{(i) MCP penalty.}
For the MCP in \eqref{eq_MCP},
the corresponding scalar proximal operator is given in closed form by
\[
\mathrm{prox}_{\rho_\eta^{\mathrm{MCP}}}^{\mathrm{scalar}}(z; \lambda) =
\begin{cases}
0, & |z| \le \lambda, \\[6pt]
\dfrac{\eta(|z| - \lambda)}{\eta - 1}\,\mathrm{sign}(z),
& \lambda < |z| \le \eta\lambda, \\[10pt]
z, & |z| > \eta\lambda.
\end{cases}
\]

\paragraph{(ii) SCAD penalty.}
For the SCAD penalty is given in \eqref{eq_SCAD_penalty},
the scalar proximal operator takes the form
\[
\mathrm{prox}_{\rho_\eta^{\mathrm{SCAD}}}^{\mathrm{scalar}}(z; \lambda) =
\begin{cases}
0, & |z| \le \lambda, \\[6pt]
\mathrm{sign}(z)\,(|z| - \lambda), &
\lambda < |z| \le 2\lambda, \\[6pt]
\mathrm{sign}(z)\,
\dfrac{(\eta - 1)|z| - \eta\lambda}{\eta - 2}, &
2\lambda < |z| \le \eta\lambda, \\[8pt]
z, & |z| > \eta\lambda.
\end{cases}
\]
Both MCP and SCAD penalties produce \emph{firm-thresholding} rules that interpolate between hard- and soft-thresholding.
For comparison, the convex ($\ell_1$-based) nuclear norm penalty corresponds to the soft-thresholding operator
\[
\mathrm{prox}_{\lambda|\cdot|}(z) = \mathrm{sign}(z)\,(|z| - \lambda)_+.
\]
In contrast, MCP and SCAD shrink small singular values toward zero while leaving large ones unpenalized, reducing estimation bias and improving rank recovery in low-rank matrix estimation.
In the proximal gradient method, these operators are applied to the singular values of the intermediate iterate.
For large matrices, truncated or randomized SVDs can be used to efficiently compute the leading singular values, making the algorithm scalable to high-dimensional settings.

Convergence of Algorithm \ref{main_algorithm} is monitored via the Frobenius norm difference 
$\|\mathbf{B}^{(t+1)} - \mathbf{B}^{(t)}\|_F < \varepsilon$,
where $\varepsilon$ is a small tolerance (e.g., $10^{-5}$).

\begin{algorithm}[H]
\caption{Proximal Gradient with Nonconvex Spectral Penalty}
\begin{algorithmic}[1]
\STATE Initialize $\mathbf{B}^{(0)} = \mathbf{0}$, choose $\alpha$, $\tau$, and $\lambda$.
\FOR{$t = 0, 1, 2, \ldots$ until convergence}
\STATE Compute residuals $\mathbf{R}^{(t)} = \mathbf{Y} - \mathbf{X}\mathbf{B}^{(t)}$.
\STATE Gradient step:
$ \displaystyle
\mathbf{G}^{(t)} = -\frac{1}{n}\mathbf{X}^\top \psi_\tau(\mathbf{R}^{(t)})
.
$
\STATE Descent update:
$
\mathbf{M}^{(t)} = \mathbf{B}^{(t)} - \alpha \mathbf{G}^{(t)}.
$
\STATE SVD: $\mathbf{M}^{(t)} = \mathbf{U}\,\mathrm{diag}(s_j)\,\mathbf{V}^\top$
\STATE Apply SCAD/MCP proximal operator:
\\
$ \qquad
s_j^{(t+1)} = \mathrm{prox}_{\text{SCAD/MCP}}(s_j; \alpha\lambda).
$
\STATE Set $\mathbf{B}^{(t+1)} = \mathbf{U}\,\mathrm{diag}(s_j^{(t+1)})\,\mathbf{V}^\top$
\ENDFOR
\end{algorithmic}
\label{main_algorithm}
\end{algorithm}

\subsection{Selection of tuning parameter $\lambda$ }

The regularization parameter \(\lambda\) and the Huber threshold \(\tau\) are jointly selected via \(K\)-fold cross-validation. A grid of candidate values is specified for both parameters, and for each pair 
\((\tau,\lambda)\), the data are randomly partitioned into \(K\) folds. For each fold 
\(k=1,\ldots,K\), the model is fitted on the training subset using the proximal gradient algorithm, yielding an estimator 
\(\widehat{\mathbf B}^{(-k)}(\tau,\lambda)\). The validation loss is then computed on the held-out fold using the squared error,
\[
\mathrm{MSE}_k(\tau,\lambda)
=
\frac{1}{n_k} \| \mathbf Y^{(k)}-\mathbf X^{(k)}\widehat{\mathbf B}^{(-k)}(\tau,\lambda) \|_F^2,
\]
where missing responses are excluded from the loss calculation. The cross-validation criterion is defined as the average validation error over all folds, and the optimal tuning parameters 
\((\widehat{\tau},\widehat{\lambda})\) are chosen to minimize this criterion.

In practice, the candidate grids for \(\lambda\) and \(\tau\) are pre-specified following data-driven heuristics. The Huber threshold \(\tau\) is selected from a small, fixed grid ({0.1, 1, 10}), which spans light to heavy truncation and is intended to capture varying degrees of robustness. For the regularization parameter 
\(\lambda\), we first compute a data-adaptive upper bound \(\lambda_{\max}\) as the largest singular value of the cross-covariance matrix between 
\(\mathbf X\) and \(\mathbf Y\), where each column is formed using only observed responses to accommodate missing data. The \(\lambda\) grid is then constructed on a logarithmic scale between 
\(\lambda_{\max}\) and a fixed lower bound 
\(\lambda_{\min}=0.05\) (default). This construction ensures that the grid covers a sufficiently wide range of regularization strengths while remaining computationally efficient.

Our method is implemented in the \texttt{R} package \texttt{rrpackrobust} available at \url{https://github.com/tienmt/rrpackrobust}.

\section{Numerical studies }
\label{sc_simulations}

\subsection{Setup}
\subsubsection*{Compared methods}
We compare our robust methods Huber$_{\rm SCAD}$, Huber$_{\rm MCP}$ against a standard non robust method rank constrain, denoted as 
`RRR' and two robust methods: a nuclear norm with Huber loss as proposed in \cite{tan2024adaptive} denoted by `Huber$_{\rm nucl}$', as well as a robust reduced rank regression via mean shift proposed in \cite{she2017robust} denoted as `R4' available from the \texttt{R} package \texttt{rrpack}. All methods use 5-fold cross validation for choosing tuning parameter.

Our codes for simulations and real data are all available on Github  (\url{https://github.com/tienmt/rrpackrobust/tree/main/code_paper_robustRRR})
to ease reader for reproducing the results.

\subsubsection*{Simulation settings}
\noindent

For given dimensionalities $n, p ,q$, we generate a covariate matrix $ X $ and a true parameter matrix $ B_0 $ as follows. The entries of $ X $ are drawn independently from a standard normal distribution, $ \mathcal{N} (0,1) $. 
The true coefficient matrix is constructed as $ B_0 = U_0 V_0^\top $, where $ U_0 \in \mathbb{R}^{p\times r} $ and $ V_0 \in \mathbb{R}^{q\times r} $ have entries generated independently from $ \mathcal{N} (0,1) $.
We consider two settings for the underlying rank of $ B_0 $ that $ r = 2$ or $ r = 5$.
The response matrix $ Y $ is then generated from model \eqref{eq_main_model} where each entries of the noise matrix drawn under a range of noise distributions to assess robustness:
\begin{itemize}
    \item Gaussian noise, $ E_{ij} \sim \mathcal{N} (0,1) $. This serves as a baseline to assess how various robust methods perform under ideal (light-tailed) conditions.
    
    \item Gaussian noise with high-variance. $ E_{ij} \sim \mathcal{N} (0,3^2) $.
    This introduces moderate heavy-tailed behavior through increased variance.
    
    \item Student noise. $ E_{ij} \sim 1.5\cdot t_3 $. This case represents heavy-tailed noise with finite variance.
    
    \item Cauchy noise. $ E_{ij} \sim Cauchy $. This reflects a more extreme heavy-tailed setting with infinite variance. 

\item Contaminated with outliers. $ E_{ij} \sim \mathcal{N} (0,1) $ or $ E_{ij} \sim t_3 $  but some portion of the observed responses are further contaminated by outliers. This scenario  evaluates robustness to contamination.
\end{itemize}

Each simulation setup is repeated 100 times and we report the average results together with its standard deviations. We report the estimation error using squared L2 norm $ \; \| \widehat{B} - B_0 \|_2^2 $ where $ \widehat{B} $ is one of the consider estimators. We also report prediction error on testing data set that are simulated in the same manner as the training data.
To measure prediction accuracy, we calculated the mean squared prediction error (MSPE) on a large, independent test set. This test set, $(X_{\text{test}}, Y_{\text{test}})$, was generated from the same model as the training data, with a fixed size of $n_{\text{test}} = 5000$. The MSPE is defined as:
$$
\text{MSPE}_{\text{test}} := \frac{1}{q\, n_{\text{test}} } 
\| Y_{{\text{test}} } - X_{{\text{test}} } \widehat{B} \|^2_2
.
$$
Finally, we assess each method's ability to correctly identify the underlying rank of the coefficient matrix.

\subsection{Simulations results}
\subsubsection*{Results with different noise}

Table \ref{tb_low_dim} summarizes the simulation results across a wide range of noise distributions, model ranks, and dimensional regimes. Overall, the proposed Huber-based methods demonstrate strong and stable performance, particularly under heavy-tailed noise, while maintaining competitiveness with classical approaches under Gaussian settings.

Under light-tailed Gaussian noise, both with variance one and variance three, all methods exhibit comparable performance in terms of estimation error and test MSPE. In these settings, the Huber estimators, R4, and RRR achieve nearly identical prediction accuracy, reflecting the fact that robustness is not essential when the model assumptions are well aligned with the data-generating mechanism. The similarity of ranks further indicates that the proposed method does not sacrifice efficiency in benign noise conditions. Among the Huber variants, the SCAD and MCP penalties consistently outperform the nuclear-norm version in estimation accuracy, suggesting that adaptive nonconvex regularization is beneficial when the true coefficient matrix is sparse or structured.

As the noise distribution becomes heavier-tailed, clear differences emerge. Under the scaled \(t_3\)
distribution, the Huber-based methods systematically outperform RRR in both estimation error and MSPE, particularly for larger rank 
\((r=5)\). This improvement highlights the advantage of local robustness induced by the Huber loss, which effectively downweights moderate outliers without explicitly modeling contamination. Compared with R4, the Huber methods generally achieve smaller estimation error and comparable or slightly better predictive performance, especially in higher-rank settings where mean-shift modeling may introduce additional variability.

The most pronounced contrast appears under Cauchy noise. In these extreme heavy-tailed scenarios, all methods experience severe degradation in MSPE, but important distinctions remain. The standard RRR estimator breaks down most severely, exhibiting extremely large estimation errors and unstable ranks. The mean-shift–based R4 method, while robust in rank recovery, suffers from substantially inflated estimation error, particularly as the rank increases. In contrast, the Huber-based estimators yield markedly smaller estimation errors than R4 and RRR across both low- and high-dimensional regimes. Although prediction error remains large for all methods, the relative stability of the Huber estimators indicates superior resistance to extreme outliers.

The high-dimensional setting \((p=120)\) reinforces these conclusions. Under Gaussian noise, all methods again perform similarly, confirming scalability of the proposed approach. Under heavy-tailed noise, especially the 
\(t_3\) and Cauchy distributions, the Huber-based methods consistently dominate RRR and often outperform R4 in estimation accuracy, while maintaining competitive rank recovery. Notably, the nuclear-norm–based Huber estimator is consistently inferior to its SCAD and MCP counterparts, underscoring the importance of nonconvex penalties in high-dimensional robust reduced-rank regression.

In summary, the simulations demonstrate that the proposed Huber method provides a favorable balance between robustness and efficiency. It matches the performance of standard RRR under light-tailed noise, substantially improves robustness under moderate and heavy-tailed distributions, and avoids the instability observed in mean-shift–based approaches under extreme contamination. These results support the use of the Huber-based framework as a reliable and broadly applicable method for robust reduced-rank regression.

The findings are further corroborated by simulation studies under a correlated design, where the predictors satisfy $ X_j \sim \mathbf{N} (0, \Sigma) $ with covariance structure $ \Sigma_{ij} = \rho_X^{|i-j|} $ and $ \rho_X = 0.5 $.
In addition, we report supplementary simulations with an increased number of responses, setting $ q = 40$. The corresponding results are summarized in Table \ref{tb_high_dim}.

\subsubsection*{Results with outliers}
Table \ref{tb_outlier} summarizes the simulation results across a wide range of data contamiated by outliers.
Under Gaussian noise with explicit contamination, the advantage of robustness becomes evident as the proportion of outliers increases. When 5\% of the observations are contaminated, the Huber-based estimators with SCAD and MCP penalties achieve the smallest estimation error and lowest test MSPE for both low and moderate ranks, while accurately recovering the true rank. 
In contrast, the standard RRR estimator already exhibits substantial degradation, with estimation error increasing by an order of magnitude relative to the robust methods. The mean-shift–based R4 approach offers partial protection against mild contamination, but its estimation error and rank recovery are noticeably less stable than those of the proposed Huber methods. The nuclear-norm–penalized Huber estimator performs considerably worse than its nonconvex counterparts, indicating limited robustness when shrinkage is overly aggressive.

As the contamination level increases to 10\% and 20\%, the differences across methods become more pronounced. 
The Huber–SCAD and Huber–MCP estimators remain the most stable in terms of both estimation accuracy and predictive performance, particularly for rank (r=2), where they consistently achieve near-optimal MSPE and correct rank recovery even under severe contamination. 
For higher rank \((r=5)\), performance naturally deteriorates for all methods, but the proposed Huber estimators continue to dominate R4 and RRR, which suffer from dramatic inflation in both estimation error and MSPE and exhibit unreliable rank estimates. Overall, these results demonstrate that the proposed Huber-based approach provides strong resilience to outliers, substantially outperforming mean-shift and non-robust alternatives as the level of contamination increases.

\subsection{Results missing data}

For the missing-data simulations, we adopt the same data-generating mechanism as before, but restrict attention to Gaussian and Student-(t) noise. After generating the complete training responses, we randomly delete 10\% or 20\% of the entries uniformly at random. As noted in the Introduction, existing robust reduced rank regression methods are not designed to accommodate missing responses; consequently, we compare our proposed approaches only with the mRRR method of \cite{luo2018leveraging}, which can handle missing data under a squared loss formulation. The results, averaged over 100 simulation replications, are reported in Table~\ref{tb_missing_Data}.

Table \ref{tb_missing_Data} demonstrates that the proposed robust method based on the Huber loss exhibits consistently strong performance across all simulation settings, particularly under heavy-tailed noise and increasing missingness. When the errors are Gaussian, the Huber–SCAD and Huber–MCP variants achieve estimation and prediction accuracy comparable to mRRR, while generally outperforming the nuclear-norm–penalized alternative, especially in terms of estimation error. Under heavy-tailed $1.5\cdot t_3$ noise, the advantage of robustness becomes pronounced: the Huber-based methods substantially reduce estimation error relative to mRRR, with competitive or improved $\text{MSPE}_{\text{test}}$, indicating enhanced stability against outliers. This pattern persists as the missing rate increases from 10\% to 20\% and across both low and moderate ranks, highlighting the resilience of the Huber loss to data contamination and incompleteness. Overall, these results confirm that incorporating the Huber loss yields a favorable bias–robustness trade-off, delivering reliable estimation and prediction in challenging missing-data and non-Gaussian settings.

\begin{table}[!ht]
\centering
\caption{\small \it Simulation for various methods under different noise conditions. The reported values are the mean across 100 simulation repetitions, with the standard deviation provided in parentheses.
$\text{MSPE}_{\text{test}}$: mean squared prediction error on testing data. }
\small
\begin{tabular}{ l l | ccc |  ccc  }
\cmidrule{1-8}
Noise & Method  
& $ \| \widehat{B} - B_0 \|_2^2 $ 
& $\text{MSPE}_{\text{test}}$
& rank
& $ \| \widehat{B} - B_0 \|_2^2 $ 
& $\text{MSPE}_{\text{test}}$
& rank
\\ \cmidrule{1-8}
\multicolumn{2}{l | }{$ n=200, p=12, q = 7 $}
&
\multicolumn{3}{c | }{$ r = 2 $} 
&
\multicolumn{3}{c}{$ r = 5 $} 
\\
\cmidrule{1-8}
$ \mathcal{N} (0,1) $  
& Huber$_{\rm SCAD}$
& 0.15 (0.00) & 1.02 (0.00) & 2.00 (0.00)
& 0.40 (0.02) & 1.05 (0.00) & 5.02 (0.14)
\\
& Huber$_{\rm MCP}$
& 0.15 (0.00) & 1.02 (0.00) & 2.00 (0.00)
& 0.40 (0.01) & 1.05 (0.00) & 5.02 (0.14)
\\
& Huber$_{\rm nucl}$
& 0.26 (0.00) & 1.03 (0.00) & 3.00 (0.00) 
& 0.43 (0.01) & 1.05 (0.00) & 5.02 (0.14) 
\\
& R4
& 0.15 (0.00) & 1.02 (0.00) & 2.00 (0.00)
& 0.40 (0.02) & 1.05 (0.00) & 5.00 (0.00) 
\\
& RRR
& 0.15 (0.00) & 1.02 (0.00) & 2.00 (0.00) 
& 0.40 (0.02) & 1.05 (0.00) & 5.00 (0.00)
\\
\cmidrule{1-8}
$ \mathcal{N} (0,3) $  
& Huber$_{\rm SCAD}$
& 1.95 (0.03) & 9.19 (0.00) & 2.02 (0.14)
& 3.37 (0.15) & 9.43 (0.04) & 5.02 (0.14)
\\
& Huber$_{\rm MCP}$
& 1.91 (0.03) & 9.19 (0.00) & 2.00 (0.00) 
& 3.35 (0.12) & 9.43 (0.04) & 5.00 (0.00) 
\\
& Huber$_{\rm nucl}$
& 2.48 (0.06) & 9.30 (0.01) & 5.00 (0.00) 
& 3.80 (0.04) & 9.50 (0.03) & 7.00 (0.00)
\\
& R4
& 1.98 (0.02) & 9.19 (0.01) & 2.00 (0.00)
& 4.19 (0.08) & 9.54 (0.03) & 4.04 (0.28)
\\
& RRR
& 1.98 (0.04) & 9.19 (0.00) & 2.00 (0.00) 
& 3.66 (0.10) & 9.48 (0.04) & 5.00 (0.00)
\\
\cmidrule{1-8}
$ 1.5 \cdot t_3 $ 
& Huber$_{\rm SCAD}$
& 0.50 (0.04) & 6.93 (0.14) & 2.00 (0.00)
& 0.85 (0.07) & 6.97 (0.12) & 4.98 (0.14)
\\
& Huber$_{\rm MCP}$
& 0.50 (0.04) & 6.93 (0.14) & 2.00 (0.00)
& 0.85 (0.07) & 6.97 (0.12) & 4.98 (0.14)
\\
& Huber$_{\rm nucl}$
& 0.74 (0.18) & 6.97 (0.12) & 2.98 (0.14)
& 0.99 (0.04) & 6.99 (0.12) & 7.00 (0.00)
\\
& R4
& 0.75 (0.02) & 6.94 (0.14) & 2.00 (0.00) 
& 1.65 (0.03) & 7.07 (0.12) & 4.00 (0.00)
\\
& RRR
& 0.69 (0.07) & 6.94 (0.13) & 2.00 (0.00)
& 1.75 (0.03) & 7.08 (0.12) & 4.04 (0.28)
\\
\cmidrule{1-8}
$ Cauchy $ 
& Huber$_{\rm SCAD}$
& 26.7 (3.69) & $>5\times 10^3$ & 6.98 (0.14)
& 7.78 (0.98) & $>5\times 10^3$ & 6.96 (0.28) 
\\
& Huber$_{\rm MCP}$
& 26.7 (3.73) & $>5\times 10^3$ & 6.94 (0.42)
& 79.6 (11.3) & $>5\times 10^3$ & 4.02 (0.14)
\\
& Huber$_{\rm nucl}$
& 25.6 (3.62) & $>5\times 10^3$ & 6.96 (0.28)
& 7.49 (0.81) & $>5\times 10^3$ & 6.96 (0.28) 
\\
& R4
& 75.5 (10.4) & $>5\times 10^3$ & 1.02 (0.14)
& 285 (40.1) & $>5\times 10^3$ & 1.08 (0.57) 
\\
& RRR
& 80.0 (0.00) & $>5\times 10^3$ & 0.00 (0.00) 
& 538 (41.4) & $>5\times 10^3$ & 0.10 (0.71)
\\ \cmidrule{1-8}
\multicolumn{2}{l | }{$ n=200, q = 7, p=120 $}
&
\multicolumn{3}{c | }{$ r = 2 $} 
&
\multicolumn{3}{c}{$ r = 5 $} 
\\
\cmidrule{1-8}
$ \mathcal{N} (0,1) $  
& Huber$_{\rm SCAD}$
& 2.98 (0.03) & 1.41 (0.00) & 2.00 (0.00)
& 7.62 (0.19) & 2.10 (0.04) & 5.00 (0.00)
\\
& Huber$_{\rm MCP}$
& 2.98 (0.03) & 1.41 (0.00) & 2.00 (0.00)
& 7.58 (0.18) & 2.09 (0.04) & 5.00 (0.00)
\\
& Huber$_{\rm nucl}$
& 6.76 (0.10) & 1.97 (0.01) & 7.00 (0.00)  
& 8.60 (0.26) & 2.24 (0.05) & 7.00 (0.00) 
\\
& R4
& 2.98 (0.03) & 1.41 (0.00) & 2.00 (0.00)
& 7.58 (0.18) & 2.09 (0.04) & 5.00 (0.00) 
\\
& RRR
& 2.98 (0.03) & 1.41 (0.00) & 2.00 (0.00) 
& 7.58 (0.18) & 2.09 (0.04) & 5.00 (0.00)
\\
\cmidrule{1-8}
$ \mathcal{N} (0,3) $  
& Huber$_{\rm SCAD}$
& 31.2 (0.52) & 13.1 (0.00) & 2.00 (0.00)
& 65.7 (0.87) & 18.2 (0.13) & 5.00 (0.00)
\\
& Huber$_{\rm MCP}$
& 31.3 (0.53) & 13.1 (0.02) & 2.00 (0.00) 
& 65.7 (0.87) & 18.2 (0.13) & 5.00 (0.00) 
\\
& Huber$_{\rm nucl}$
& 53.0 (0.87) & 16.5 (0.09) & 7.00 (0.00) 
& 79.1 (0.71) & 20.3 (0.12) & 7.00 (0.00)
\\
& R4
& 31.1 (0.54) & 13.1 (0.00) & 2.00 (0.00)
& 65.7 (0.87) & 18.2 (0.13) & 5.00 (0.00)
\\
& RRR
& 31.1 (0.54) & 13.1 (0.00) & 2.00 (0.00) 
& 65.7 (0.87) & 18.2 (0.13) & 5.00 (0.00)
\\
\cmidrule{1-8}
$ 1.5 \cdot t_3 $ 
& Huber$_{\rm SCAD}$
& 14.1 (0.30) & 9.94 (0.28) & 2.00 (0.00)
& 35.7 (0.74) & 12.9 (0.07) & 5.00 (0.00) 
\\
& Huber$_{\rm MCP}$
& 14.1 (0.30) & 9.94 (0.28) & 2.00 (0.00)
& 35.7 (0.74) & 12.9 (0.07) & 5.00 (0.00)
\\
& Huber$_{\rm nucl}$
& 30.3 (0.38) & 12.2 (0.28) & 7.00 (0.00)
& 41.2 (0.26) & 13.6 (0.00) & 7.00 (0.00)
\\
& R4
& 20.5 (0.33) & 10.8 (0.29) & 2.00 (0.00) 
& 41.1 (0.20) & 13.6 (0.04) & 5.00 (0.00)
\\
& RRR
& 29.3 (1.82) & 12.2 (0.51) & 2.00 (0.00)
& 55.9 (2.01) & 16.1 (0.30) & 5.00 (0.00)
\\
\cmidrule{1-8}
$ Cauchy $ 
& Huber$_{\rm SCAD}$
& 15.2 (0.18) & $>5\times 10^3$ & 2.00 (0.00) 
& 85.0 (0.40) & $>5\times 10^3$ & 5.00 (0.00) 
\\
& Huber$_{\rm MCP}$
& 15.2 (0.18) & $>5\times 10^3$ & 2.00 (0.00)
& 85.0 (0.40) & $>5\times 10^3$ & 5.00 (0.00)
\\
& Huber$_{\rm nucl}$
& 43.1 (1.31) & $>5\times 10^3$ & 6.98 (0.14)
& 118 (2.04) & $>5\times 10^3$ & 7.00 (0.00)
\\
& R4
& 257 (26.6) & $>5\times 10^3$ & 1.98 (0.14) 
& 558 (37.2) & $>5\times 10^3$ & 4.96 (0.28) 
\\
& RRR
& 1279 (0.00) & $>5\times 10^3$ & 0.00 (0.00) 
& 4689 (0.00) & $>5\times 10^3$ & 0.00 (0.00)
\\
\cmidrule{1-8}
\end{tabular}
\label{tb_low_dim}
\end{table}

\begin{table}[!ht]
\centering
\caption{\small \it Simulation results for multiple methods under a correlated design matrix ($\rho_X = 0.5 $) and with an increased number of responses $ q = 40$. The reported values are the mean across 100 simulation repetitions, with the standard deviation provided in parentheses. 
$\text{MSPE}_{\text{test}}$: mean squared prediction error on testing data. }
\small
\begin{tabular}{ l l | ccc | c cccc }
		\hline 
Noise & Method  
& $ \| \widehat{B} - B_0 \|_2^2 $ 
& $\text{MSPE}_{\text{test}}$
& rank
& $ \| \widehat{B} - B_0 \|_2^2 $ 
& $\text{MSPE}_{\text{test}}$
& rank
\\
\hline
\multicolumn{8}{l}{$\bf{n=200, p=12, q = 7, \rho_X = 0.5 }$} 
\\
& &
\multicolumn{3}{c | }{$ r = 2 $} 
&
\multicolumn{3}{c}{$ r = 5 $} 
\\
\cmidrule{1-5} \cmidrule{6-8} 
$ \mathcal{N} (0,1) $  
& Huber$_{\rm SCAD}$
& 0.27 (0.00) & 1.03 (0.00) & 2.00 (0.00) 
& 0.57 (0.02) & 1.05 (0.00) & 5.00 (0.00)
\\
& Huber$_{\rm MCP}$
& 0.27 (0.00) & 1.03 (0.00) & 2.00 (0.00) 
& 0.58 (0.02) & 1.05 (0.00) & 5.00 (0.00)
\\
& Huber$_{\rm nucl}$
& 0.37 (0.00) & 1.04 (0.00) & 2.00 (0.00)
& 0.70 (0.00) & 1.06 (0.00) & 6.00 (0.00)
\\
& R4
& 0.27 (0.00) & 1.03 (0.00) & 2.00 (0.00)
& 0.58 (0.03) & 1.05 (0.00) & 5.00 (0.00)
\\
& RRR
& 0.27 (0.00) & 1.03 (0.00) & 2.00 (0.00) 
& 0.57 (0.02) & 1.05 (0.00) & 5.00 (0.00) 
\\
		\hline
$ \mathcal{N} (0,3) $  
& Huber$_{\rm SCAD}$
& 2.51 (0.07) & 9.23 (0.00) & 2.00 (0.00)
& 5.40 (0.13) & 9.48 (0.01) & 5.98 (0.14)
\\
& Huber$_{\rm MCP}$
& 2.51 (0.07) & 9.23 (0.00) & 2.00 (0.00)
& 5.23 (0.16) & 9.48 (0.01) & 5.00 (0.00)
\\
& Huber$_{\rm nucl}$
& 5.27 (0.09) & 9.42 (0.01) & 5.00 (0.00)
& 5.92 (0.00) & 9.52 (0.01) & 7.00 (0.00)
\\
& R4
& 2.53 (0.08) & 9.23 (0.01) & 2.00 (0.00) 
& 5.34 (0.20) & 9.49 (0.01) & 5.00 (0.00)
\\
& RRR
& 2.51 (0.07) & 9.23 (0.00) & 2.00 (0.00)
& 5.23 (0.16) & 9.48 (0.01) & 5.00 (0.00)
\\
		\hline
$ t_3 $ 
& Huber$_{\rm SCAD}$
& 0.57 (0.03) & 6.91 (0.07) & 2.00 (0.00)
& 1.36 (0.21) & 7.00 (0.03) & 5.00 (0.00)
\\
& Huber$_{\rm MCP}$
& 0.57 (0.03) & 6.91 (0.07) & 2.00 (0.00)
& 1.36 (0.21) & 7.00 (0.03) & 5.00 (0.00)
\\
& Huber$_{\rm nucl}$
& 1.05 (0.02) & 6.95 (0.07) & 5.00 (0.00)
& 1.61 (0.19) & 7.03 (0.03) & 6.00 (0.00)
\\
& R4
& 1.20 (0.04) & 6.96 (0.07) & 2.00 (0.00)
& 2.45 (0.23) & 7.08 (0.03) & 5.00 (0.00)
\\
& RRR
& 1.41 (0.02) & 6.98 (0.07) & 2.00 (0.00)
& 3.24 (0.09) & 7.13 (0.02) & 5.00 (0.00)
\\
		\hline
$ Cauchy $ 
& Huber$_{\rm SCAD}$
& 3.31 (0.12) & $>5\times 10^3$ & 2.00 (0.00) 
& 16.4 (33.1) & $>5\times 10^3$ & 6.86 (0.99) 
\\
& Huber$_{\rm MCP}$
& 3.31 (0.12) & $>5\times 10^3$ & 2.00 (0.00) 
& 14.3 (33.4) & $>5\times 10^3$ & 4.90 (0.71) 
\\
& Huber$_{\rm nucl}$
& 10.0 (1.29) & $>5\times 10^3$ & 6.92 (0.57) 
& 15.7 (33.2) & $>5\times 10^3$ & 6.86 (0.99)
\\
& R4
& 107 (1.71) & $>5\times 10^3$ & 1.00 (0.00)
& 163 (0.99) & $>5\times 10^3$ & 1.00 (0.00) 
\\
& RRR
& 217 (0.00) & $>5\times 10^3$ & 0.00 (0.00) 
& 246 (0.00) & $>5\times 10^3$ & 0.00 (0.00) 
\\ \hline
\multicolumn{3}{l}{$\bf{n=200, p= 40 , q = 7 }$} 
\\
\cmidrule{1-5} \cmidrule{6-8} 
$ \mathcal{N} (0,1) $  
& Huber$_{\rm SCAD}$
& 0.59 (0.00) & 1.03 (0.00) & 2.00 (0.00) 
& 1.22 (0.02) & 1.04 (0.00) & 5.00 (0.00) 
\\
& Huber$_{\rm MCP}$
& 0.59 (0.00) & 1.03 (0.00) & 2.00 (0.00)
& 1.22 (0.02) & 1.04 (0.00) & 5.00 (0.00)
\\
& Huber$_{\rm nucl}$
& 1.23 (0.02) & 1.04 (0.00) & 2.00 (0.00)
& 1.79 (0.04) & 1.06 (0.00) & 6.02 (0.14) 
\\
& R4
& 0.59 (0.01) & 1.03 (0.00) & 2.00 (0.00)
& 1.22 (0.02) & 1.04 (0.00) & 5.00 (0.00) 
\\
& RRR
& 0.59 (0.00) & 1.03 (0.00) & 2.00 (0.00)
& 1.22 (0.02) & 1.04 (0.00) & 5.00 (0.00)
\\
		\hline
$ \mathcal{N} (0,3) $  
& Huber$_{\rm SCAD}$
& 5.33 (0.14) & 9.26 (0.02) & 2.00 (0.00)
& 10.9 (0.15) & 9.38 (0.00) & 5.00 (0.00)
\\
& Huber$_{\rm MCP}$
& 5.33 (0.14) & 9.26 (0.02) & 2.00 (0.00)
& 10.9 (0.15) & 9.38 (0.00) & 5.00 (0.00)
\\
& Huber$_{\rm nucl}$
& 11.5 (0.48) & 9.41 (0.03) & 3.00 (0.00) 
& 16.1 (0.33) & 9.51 (0.00) & 8.02 (0.14)
\\
& R4
& 5.37 (0.14) & 9.26 (0.02) & 2.00 (0.00) 
& 11.0 (0.13) & 9.38 (0.00) & 5.00 (0.00) 
\\
& RRR
& 5.33 (0.14) & 9.26 (0.02) & 2.00 (0.00)
& 10.9 (0.14) & 9.38 (0.00) & 5.00 (0.00) 
\\
		\hline
$ t_3 $ 
& Huber$_{\rm SCAD}$
& 1.63 (0.04) & 6.50 (0.02) & 2.00 (0.00)
& 4.69 (0.03) & 6.57 (0.01) & 5.00 (0.00)
\\
& Huber$_{\rm MCP}$
& 1.63 (0.04) & 6.50 (0.02) & 2.00 (0.00)
& 4.69 (0.03) & 6.57 (0.01) & 5.00 (0.00)
\\
& Huber$_{\rm nucl}$
& 4.54 (0.06) & 6.57 (0.02) & 2.98 (0.14) 
& 6.63 (0.02) & 6.62 (0.01) & 8.02 (0.14)
\\
& R4
& 2.90 (0.14) & 6.53 (0.02) & 2.00 (0.00) 
& 7.35 (0.08) & 6.65 (0.01) & 5.00 (0.00)
\\
& RRR
& 3.06 (0.09) & 6.54 (0.02) & 2.00 (0.00)
& 7.42 (0.10) & 6.65 (0.01) & 5.00 (0.00) 
\\
		\hline
$ Cauchy $ 
& Huber$_{\rm SCAD}$
& 18.3 (2.47) & $> 10^4 $ & 8.86 (0.99)
& 26.8 (0.50) & $> 10^4 $ & 5.00 (0.00)
\\
& Huber$_{\rm MCP}$
& 7.17 (0.86) & $> 10^4 $ & 11.8 (1.41) 
& 28.4 (3.61) & $> 10^4 $ & 10.8 (0.85)
\\
& Huber$_{\rm nucl}$
& 15.2 (1.89) & $> 10^4 $ & 8.92 (0.57)
& 108 (14.2) & $> 10^4 $ & 5.00 (0.00)
\\
& R4
& 196 (2.36) & $> 10^4 $ & 1.00 (0.00)
& 578 (96.3) & $> 10^4 $ & 2.96 (0.28)
\\
& RRR
& 862 (0.00) & $> 10^4 $ & 0.00 (0.00)
& 2039 (0.00) & $> 10^4 $ & 0.00 (0.00) 
\\
		\hline
\end{tabular}
\label{tb_high_dim}
\end{table}

\begin{table}[!ht]
\centering
\caption{\small \it Simulation results  under the setting $ p = 12, q = 7, n = 200 $ with the proportion of outliers increased to 5\%, 10\% and 20\%.
 The reported values are the mean across 100 simulation repetitions, with the standard deviation provided in parentheses.
 $\text{MSPE}_{\text{test}}$: mean squared prediction error on testing data.  }
\small
\begin{tabular}{ l l | ccc | ccc }
 \cmidrule{1-8}
Noise & Method  
& $ \| \widehat{B} - B_0 \|_2^2 $ 
& $\text{MSPE}_{\text{test}}$
& rank
& $ \| \widehat{B} - B_0 \|_2^2 $ 
& $\text{MSPE}_{\text{test}}$
& rank
\\
 \cmidrule{1-8}
& &
\multicolumn{3}{c | }{$ r = 2 $} 
&
\multicolumn{3}{c}{$ r = 5 $} 
\\
 \cmidrule{1-8}
$ \mathcal{N} (0,1), $  
& Huber$_{\rm SCAD}$
& 0.41 (0.01) & 1.05 (0.00) & 2.00 (0.00)
& 1.41 (0.10) & 1.18 (0.01) & 5.00 (0.00)
\\
5\% 
 & Huber$_{\rm MCP}$
& 0.35 (0.02) & 1.05 (0.00) & 2.00 (0.00)
& 1.41 (0.10) & 1.18 (0.01) & 5.00 (0.00)
\\
outliers
& Huber$_{\rm nucl}$
& 1.45 (0.07) & 7.07 (0.01) & 6.90 (0.71)
& 1.66 (0.21) & 7.11 (0.04) & 6.98 (0.14) 
\\
& R4
& 0.78 (0.05) & 1.10 (0.00) & 2.98 (0.14) 
& 1.85 (0.12) & 1.25 (0.02) & 5.00 (0.00)
\\
& RRR
& 22.0 (0.66) & 4.17 (0.10) & 2.98 (0.14)
& 46.6 (3.51) & 7.69 (0.48) & 5.94 (0.42)
\\
 \cmidrule{1-8}
$ \mathcal{N} (0,1), $  
& Huber$_{\rm SCAD}$
& 0.23 (0.05) & 1.04 (0.01) & 2.00 (0.00)
& 16.8 (0.35) & 3.42 (0.05) & 4.02 (0.14)
\\
10\% 
 & Huber$_{\rm MCP}$
& 0.23 (0.05) & 1.04 (0.01) & 2.00 (0.00)
& 16.8 (0.35) & 3.42 (0.05) & 4.02 (0.14)
\\
outliers
& Huber$_{\rm nucl}$
& 2.34 (0.02) & 1.32 (0.00) & 2.00 (0.00)   
& 33.4 (4.11) & 11.8 (0.59) & 6.00 (0.00)
\\
& R4
& 104 (9.49) & 15.4 (1.36) & 1.04 (0.28) 
& 35.1 (1.00) & 5.92 (0.12) & 4.02 (0.14)
\\
& RRR
& 114 (0.84) & 16.7 (0.35) & 1.04 (0.28)
& 99.2 (0.38) & 15.5 (0.15) & 6.02 (0.14)
\\
 \cmidrule{1-8}
$ \mathcal{N} (0,1), $  
& Huber$_{\rm SCAD}$
& 21.8 (1.54) & 4.26 (0.24) & 2.00 (0.00) 
& 58.5 (4.03) & 9.62 (0.61) & 5.00 (0.00)
\\
20\% 
 & Huber$_{\rm MCP}$
& 21.8 (1.54) & 4.27 (0.24) & 2.00 (0.00) 
& 58.7 (4.02) & 9.65 (0.60) & 5.00 (0.00)  
\\
outliers
& Huber$_{\rm nucl}$
& 19.9 (0.07) & 3.97 (0.04) & 2.00 (0.00)
& 107 (5.84) & 17.2 (0.90) & 5.00 (0.00) 
\\
& R4
& 35.5 (0.15) & 6.33 (0.07) & 1.00 (0.00)
& 184 (5.25) & 28.2 (0.84) & 2.00 (0.00)
\\
& RRR
& 35.7 (6.36) & 6.37 (0.88) & 1.02 (0.14) 
& 160 (11.4) & 24.2 (1.65) & 4.96 (0.28) 
\\
 \cmidrule{1-8}
 \hline
$ t_3 $  
& Huber$_{\rm SCAD}$
& 0.71 (0.00) & 6.05 (0.01) & 2.00 (0.00)
& 2.08 (0.12) & 6.68 (0.01) & 5.00 (0.00)
\\
5\% 
 & Huber$_{\rm MCP}$
& 0.71 (0.00) & 6.05 (0.01) & 2.00 (0.00)
& 2.08 (0.12) & 6.68 (0.01) & 5.00 (0.00)
\\
outliers
& Huber$_{\rm nucl}$
& 1.91 (0.13) & 6.64 (0.08) & 3.02 (0.14)
& 2.64 (0.16) & 6.74 (0.01) & 6.04 (0.28)
\\
& R4
& 2.66 (0.12) & 6.37 (0.01) & 2.00 (0.00)
& 4.15 (0.01) & 6.89 (0.00) & 5.00 (0.00)
\\
& RRR
& 22.1 (0.98) & 9.13 (0.15) & 2.96 (0.28) 
& 19.6 (0.59) & 8.30 (0.11) & 5.00 (0.00)
\\
 \cmidrule{1-8}
$ t_3 $  
& Huber$_{\rm SCAD}$
& 1.32 (0.01) & 6.16 (0.32) & 2.00 (0.00)
& 6.77 (0.39) & 6.92 (0.05) & 4.00 (0.00)
\\
10\% 
 & Huber$_{\rm MCP}$
& 1.32 (0.01) & 6.16 (0.32) & 2.00 (0.00)
& 6.74 (0.61) & 6.92 (0.08) & 4.02 (0.14) 
\\
outliers
& Huber$_{\rm nucl}$
& 33.1 (4.37) & 10.4 (0.97) & 2.00 (0.00) 
& 10.3 (0.10) & 7.34 (0.04) & 4.00 (0.00) 
\\
& R4
& 12.1 (7.82) & 7.77 (1.48) & 1.98 (0.14)
& 43.3 (1.98) & 12.1 (0.26) & 3.02 (0.14) 
\\
& RRR
& 16.3 (7.58) & 8.34 (1.45) & 1.98 (0.14)
& 40.8 (0.31) & 11.9 (0.00) & 5.02 (0.14)
\\
 \cmidrule{1-8}
$ t_3 $  
& Huber$_{\rm SCAD}$
& 62.1 (7.03) & 14.9 (0.90) & 2.00 (0.00)
& 82.0 (8.78) & 17.9 (1.30) & 4.00 (0.00)
\\
20\% 
& Huber$_{\rm MCP}$
& 62.2 (7.03) & 14.9 (0.89) & 2.00 (0.00)
& 84.4 (9.19) & 18.2 (1.35) & 4.00 (0.00)
\\
outliers
& Huber$_{\rm nucl}$
& 27.7 (0.80) & 9.94 (0.01) & 2.00 (0.00) 
& 22.7 (1.05) & 9.12 (0.14) & 4.00 (0.00)  
\\
& R4
& 87.0 (1.71) & 19.2 (0.23) & 1.00 (0.00)
& 77.0 (2.83) & 16.6 (0.37) & 2.02 (0.14) 
\\
& RRR
& 51.2 (4.46) & 13.6 (0.58) & 2.00 (0.00)
& 71.3 (7.42) & 16.1 (0.94) & 3.06 (0.42) 
\\
		\hline
		\hline	
\end{tabular}
\label{tb_outlier}
\end{table}

\begin{table}[!ht]
\centering
\caption{\small \it Simulation for missing data $ n=200, p=12, q = 7 $ with missing data. The reported values are the mean across 100 simulation repetitions, with the standard deviation provided in parentheses. 
$\text{MSPE}_{\text{test}}$: mean squared prediction error on testing data. }
\small
\begin{tabular}{ l l | ccc |  ccc  }
\cmidrule{1-8}
Noise & Method  
& $ \| \widehat{B} - B_0 \|_2^2 $ 
& $\text{MSPE}_{\text{test}}$
& rank
& $ \| \widehat{B} - B_0 \|_2^2 $ 
& $\text{MSPE}_{\text{test}}$
& rank
\\ \cmidrule{1-8}
\multicolumn{2}{c | }{10\% missing}
&
\multicolumn{3}{c | }{$ r = 2 $} 
&
\multicolumn{3}{c}{$ r = 5 $} 
\\
\cmidrule{1-8}
$ \mathcal{N} (0,1) $  
& Huber$_{\rm SCAD}$
& 0.21 (0.00) & 0.98 (0.02) & 2.00 (0.00) 
& 0.45 (0.02) & 0.98 (0.02) & 5.00 (0.00) 
\\
& Huber$_{\rm MCP}$
& 0.21 (0.00) & 0.98 (0.02) & 2.00 (0.00) 
& 0.48 (0.02) & 0.99 (0.02) & 5.00 (0.00) 
\\
& Huber$_{\rm nucl}$
& 0.44 (0.02) & 0.98 (0.02) & 2.98 (0.14) 
& 0.52 (0.02) & 0.99 (0.01) & 5.00 (0.00)  
\\
& mRRR
& 0.21 (0.00) & 0.98 (0.02) & 2.00 (0.00) 
& 0.44 (0.02) & 0.98 (0.01) & 5.00 (0.00) 
\\
\cmidrule{1-8}
$ 1.5 \cdot t_3 $ 
& Huber$_{\rm SCAD}$
& 0.53 (0.02) & 6.35 (0.26) & 2.00 (0.00)
& 1.12 (0.05) & 6.44 (0.05) & 5.00 (0.00)
\\
& Huber$_{\rm MCP}$
& 0.53 (0.02) & 6.35 (0.26) & 2.00 (0.00)
& 1.12 (0.05) & 6.44 (0.05) & 5.00 (0.00)
\\
& Huber$_{\rm nucl}$
& 0.66 (0.04) & 6.42 (0.25) & 3.00 (0.00) 
& 1.26 (0.07) & 6.54 (0.06) & 5.98 (0.14) 
\\
& mRRR
& 1.26 (0.02) & 6.38 (0.24) & 2.00 (0.00)
& 2.21 (0.03) & 6.62 (0.04) & 5.00 (0.00)
\\ \cmidrule{1-8}
\multicolumn{2}{c | }{20\% missing}
&
\multicolumn{3}{c | }{$ r = 2 $} 
&
\multicolumn{3}{c}{$ r = 5 $} 
\\
\cmidrule{1-8}
$ \mathcal{N} (0,1) $  
& Huber$_{\rm SCAD}$
& 0.29 (0.01) & 0.92 (0.02) & 2.00 (0.00) 
& 0.44 (0.00) & 0.91 (0.02) & 5.00 (0.00) 
\\
& Huber$_{\rm MCP}$
& 0.29 (0.01) & 0.92 (0.02) & 2.00 (0.00) 
& 0.43 (0.00) & 0.91 (0.02) & 5.00 (0.00) 
\\
& Huber$_{\rm nucl}$
& 0.38 (0.02) & 0.90 (0.02) & 3.00 (0.00)  
& 0.56 (0.00) & 0.93 (0.02) & 6.00 (0.00) 
\\
& mRRR
& 0.28 (0.01) & 0.92 (0.02) & 2.00 (0.00) 
& 0.43 (0.00) & 0.91 (0.02) & 5.00 (0.00) 
\\
\cmidrule{1-8}
$ 1.5 \cdot t_3 $ 
& Huber$_{\rm SCAD}$
& 0.78 (0.01) & 6.24 (0.04) & 2.00 (0.00) 
& 1.18 (0.02) & 6.28 (0.01) & 5.00 (0.00) 
\\
& Huber$_{\rm MCP}$
& 0.78 (0.01) & 6.24 (0.04) & 2.00 (0.00) 
& 1.18 (0.02) & 6.28 (0.01) & 5.00 (0.00)
\\
& Huber$_{\rm nucl}$
& 1.26 (0.05) & 6.39 (0.06) & 2.02 (0.14) 
& 1.39 (0.04) & 6.35 (0.01) & 7.00 (0.00) 
\\
& mRRR
& 1.03 (0.07) & 6.20 (0.01) & 2.00 (0.00) 
& 2.10 (0.02) & 6.36 (0.02) & 5.00 (0.00) 
\\
\cmidrule{1-8}
\end{tabular}
\label{tb_missing_Data}
\end{table}

\section{Application to cancer cell line data set}
\label{sc_application}

We analyze the NCI-60 cancer cell line data set, which consists of gene expression measurements and protein expression profiles collected on \(n = 59\) cell lines. The predictor matrix
contains 22283 standardized gene expression levels, while the multivariate response matrix
records 162 standardized protein abundances. Both \(X\) and \(Y\) were column-centered and scaled to unit variance prior to analysis. The data is publicly available in the \texttt{R} package \texttt{robustHD} \citep{alfons2021robusthd}.

To construct the training and testing datasets, we proceeded as follows. The multivariate response matrix \(Y\) was defined using the first 20 protein measurements.
To reduce dimensionality and focus on the most relevant predictors, we computed the empirical cross-covariance matrix between \(X\) and \(Y\) and derived an aggregate correlation score for each predictor based on the \(\ell_2\)-norm of its cross-covariance vector across responses. 
The 100 predictors with the largest scores were retained, yielding a reduced design matrix.
Model evaluation was then performed via repeated random data splitting: in each of 100 repetitions, 9 observations were randomly selected without replacement to form the test set, while the remaining observations constituted the training set. 
Prediction accuracy and rank estimation were subsequently assessed by averaging results across these repeated train–test splits.

Table \ref{tb_realdata} summarizes the performance of the proposed robust reduced-rank estimators on the real data across varying levels of missingness. 
Overall, the Huber-based methods with nonconvex spectral penalties—Huber$_{\rm MCP}$ and Huber$_{\rm SCAD}$—consistently achieve the lowest prediction errors, substantially outperforming the classical mRRR and the robust R4 method.
This gain is already evident in the fully observed setting and persists as the proportion of missing responses increases, indicating strong robustness to both heavy-tailed noise and incomplete data. In terms of rank recovery, 
the nonconvex penalties yield stable and interpretable low-rank estimates: Huber$_{\rm MCP}$ tends to select more parsimonious ranks, 
while Huber$_{\rm SCAD}$ produces slightly higher but still stable rank estimates, reflecting their differing bias–variance trade-offs. 
By contrast, mRRR exhibits unstable rank behavior under missingness, often collapsing to degenerate solutions, and R4 fails to adapt.
The Huber nuclear-norm estimator performs comparably in prediction but typically selects higher ranks, highlighting the advantage of nonconvex spectral regularization for simultaneous robustness and effective rank selection in real-data settings with missing responses.

\begin{table}[!ht]
\centering
\caption{\small \it Results for NCI-60 cancer cell line data. The reported values are the mean across 100 repetitions, with the standard deviation provided in parentheses. mRRR: method for handling missing data with squared loss. R4: a robust reduced rank regression method via mean shift and does not work with missing data. }
\footnotesize
\begin{tabular}{  l | cccc ccccc }
		\hline \hline
 & mRRR & R4 & Huber$_{\rm MCP}$ & Huber$_{\rm SCAD}$ & Huber$_{\rm nucl}$ 
 \\ \hline
prediction 
& 0.71 (0.03) & 1.98 (0.48) & 0.49 (0.05) & 0.49 (0.04) & 0.49 (0.04) 
\\
rank estimate 
& 1.93 (1.34)  & 20.0 (0.00) &  4.00 (0.00) & 5.99 (0.10) & 5.99  (0.10) 
 \\ \hline
\multicolumn{6}{c  }{ missing 10\% }  
  \\ \hline
prediction 
& 0.71 (0.03) & \_\_ & 0.49 (0.04) & 0.50 (0.04) & 0.50 (0.04) 
\\
rank estimate 
& 0.00 (0.00)  & \_\_ & 4.00 (0.00) & 5.98 (0.20) & 6.02 (0.20)  
 \\ \hline
\multicolumn{6}{c  }{ missing 20\% }  
 \\ 
  \hline
prediction 
& 0.70 (0.00) & \_\_ & 0.50 (0.02) & 0.51 (0.02) & 0.51 (0.02) 
\\
rank estimate 
& 0.01 (0.10) & \_\_ & 3.01 (0.10) & 5.01 (0.10) & 5.01 (0.10)  
 \\ \hline
\multicolumn{6}{c  }{ missing 30\% }  
\\ 
  \hline
prediction 
& 0.55 (0.04) & \_\_ & 0.52 (0.03) & 0.52 (0.03) & 0.52 (0.03) 
\\
rank estimate 
& 2.98 (0.20) & \_\_ & 3.99 (0.10) & 3.99 (0.10) & 4.01 (0.10)  
\\
		\hline	
\end{tabular}
\label{tb_realdata}
\end{table}

\section{Conclusion and Discussion}
\label{sc_concluson}

In this paper, we proposed a robust reduced rank regression framework that simultaneously addresses heavy-tailed noise, outliers, and missing responses in multivariate regression. By combining the Huber loss with nonconvex spectral penalties, specifically MCP and SCAD, our approach effectively balances robustness and low-rank structure recovery. The proposed formulation extends existing RRR methodologies by accommodating incomplete response matrices without imputation, while mitigating the estimation bias commonly induced by convex nuclear-norm regularization.

From a computational perspective, we developed an efficient proximal gradient algorithm tailored to the proposed nonconvex objective. The algorithm leverages the Lipschitz continuity of the Huber loss gradient and closed-form proximal mappings of the MCP and SCAD penalties, resulting in scalable updates based on singular value thresholding. This structure enables practical implementation in high-dimensional settings through truncated or randomized SVDs, while maintaining numerical stability despite the inherent nonconvexity of the problem.

Extensive simulation studies and real data analyses demonstrate the advantages of the proposed methods over existing alternatives. In particular, our approach consistently improves estimation accuracy, rank recovery, and predictive performance under heavy-tailed noise and increasing levels of missingness, outperforming both nuclear-norm–based robust methods and non-robust RRR estimators. These results suggest that nonconvex spectral regularization combined with robust loss functions provides a powerful and flexible tool for multivariate regression in challenging data. They opens several avenues for future work, including theoretical analysis under broader missingness mechanisms and extensions to other robust loss functions. Extensions to Bayesian inference are the objective of future research directions \citep{mai2025hightobit,mai2024concentration,mai2026optimal,mai2025bayesian,mai2025handling,mai2025sparse}.

\subsection*{Acknowledgments}
The views, results, and opinions presented in this paper are solely those of the author and do not, in any form, represent those of the Norwegian Institute of Public Health.

\subsection*{Conflicts of interest/Competing interests}
The author declares no potential conflict of interests.


\begin{thebibliography}{}
	
	\bibitem[Alfons, 2021]{alfons2021robusthd}
	Alfons, A. (2021).
	\newblock {robustHD: An R package for robust regression with high-dimensional
		data}.
	\newblock {\em Journal of Open Source Software}, 6(67):3786.
	
	\bibitem[Anderson, 1951]{anderson1951estimating}
	Anderson, T.~W. (1951).
	\newblock Estimating linear restrictions on regression coefficients for
	multivariate normal distributions.
	\newblock {\em Annals of mathematical statistics}, 22(3):327--351.
	
	\bibitem[Bing and Wegkamp, 2019]{bing2019adaptive}
	Bing, X. and Wegkamp, M.~H. (2019).
	\newblock Adaptive estimation of the rank of the coefficient matrix in
	high-dimensional multivariate response regression models.
	\newblock {\em The Annals of Statistics}, 47(6):3157--3184.
	
	\bibitem[Bunea et~al., 2011]{bunea2011optimal}
	Bunea, F., She, Y., and Wegkamp, M.~H. (2011).
	\newblock Optimal selection of reduced rank estimators of high-dimensional
	matrices.
	\newblock {\em The Annals of Statistics}, 39(2):1282--1309.
	
	\bibitem[Bunea et~al., 2012]{bunea2012joint}
	Bunea, F., She, Y., and Wegkamp, M.~H. (2012).
	\newblock Joint variable and rank selection for parsimonious estimation of
	high-dimensional matrices.
	\newblock {\em The Annals of Statistics}, 40(5):2359--2388.
	
	\bibitem[Chakraborty et~al., 2020]{chakraborty2020bayesian}
	Chakraborty, A., Bhattacharya, A., and Mallick, B.~K. (2020).
	\newblock {B}ayesian sparse multiple regression for simultaneous rank reduction
	and variable selection.
	\newblock {\em Biometrika}, 107(1):205--221.
	
	\bibitem[Chen et~al., 2013]{chen2013reduced}
	Chen, K., Dong, H., and Chan, K.-S. (2013).
	\newblock Reduced rank regression via adaptive nuclear norm penalization.
	\newblock {\em Biometrika}, 100(4):901--920.
	
	\bibitem[Cook, 2018]{cook2018introduction}
	Cook, R.~D. (2018).
	\newblock {\em An introduction to envelopes: dimension reduction for efficient
		estimation in multivariate statistics}.
	\newblock John Wiley \& Sons.
	
	\bibitem[Fan and Li, 2001]{fan2001variable}
	Fan, J. and Li, R. (2001).
	\newblock Variable selection via nonconcave penalized likelihood and its oracle
	properties.
	\newblock {\em Journal of the American statistical Association},
	96(456):1348--1360.
	
	\bibitem[Geweke, 1996]{geweke1996bayesian}
	Geweke, J. (1996).
	\newblock {B}ayesian reduced rank regression in econometrics.
	\newblock {\em Journal of econometrics}, 75(1):121--146.
	
	\bibitem[Giraud, 2021]{giraud2021introduction}
	Giraud, C. (2021).
	\newblock {\em Introduction to high-dimensional statistics}.
	\newblock Chapman and Hall/CRC.
	
	\bibitem[Goh et~al., 2017]{goh2017bayesian}
	Goh, G., Dey, D.~K., and Chen, K. (2017).
	\newblock {B}ayesian sparse reduced rank multivariate regression.
	\newblock {\em Journal of multivariate analysis}, 157:14--28.
	
	\bibitem[Huber, 1964]{huber1964robust}
	Huber, P.~J. (1964).
	\newblock Robust estimation of a location parameter.
	\newblock {\em The Annals of Mathematical Statistics}, 35(1):73--101.
	
	\bibitem[Huber, 1996]{huber1996robust}
	Huber, P.~J. (1996).
	\newblock {\em Robust statistical procedures}.
	\newblock SIAM.
	
	\bibitem[Izenman, 1975]{izenman1975reduced}
	Izenman, A.~J. (1975).
	\newblock Reduced-rank regression for the multivariate linear model.
	\newblock {\em Journal of multivariate analysis}, 5(2):248--264.
	
	\bibitem[Izenman, 2008]{izenman2008modern}
	Izenman, A.~J. (2008).
	\newblock Modern multivariate statistical techniques: Regression,
	classification and manifold learning.
	\newblock {\em Springer Texts in Statistics}, 10:978.
	
	\bibitem[Luo et~al., 2018]{luo2018leveraging}
	Luo, C., Liang, J., Li, G., Wang, F., Zhang, C., Dey, D.~K., and Chen, K.
	(2018).
	\newblock Leveraging mixed and incomplete outcomes via reduced-rank modeling.
	\newblock {\em Journal of Multivariate Analysis}, 167:378--394.
	
	\bibitem[Ma et~al., 2024]{ma2024d4r}
	Ma, X., Wei, L., and Liang, W. (2024).
	\newblock D4r: Doubly robust reduced rank regression in high dimension.
	\newblock {\em Journal of Statistical Planning and Inference}, 232:106162.
	
	\bibitem[Mai, 2023a]{mai2023bilinear}
	Mai, T.~T. (2023a).
	\newblock From bilinear regression to inductive matrix completion: a
	quasi-bayesian analysis.
	\newblock {\em Entropy}, 25(2):333.
	
	\bibitem[Mai, 2023b]{mai2023reduced}
	Mai, T.~T. (2023b).
	\newblock A reduced-rank approach to predicting multiple binary responses
	through machine learning.
	\newblock {\em Statistics and Computing}, 33(6):136.
	
	\bibitem[Mai, 2024]{mai2024concentration}
	Mai, T.~T. (2024).
	\newblock Concentration of a sparse bayesian model with horseshoe prior in
	estimating high-dimensional precision matrix.
	\newblock {\em Stat}, 13(4):e70008.
	
	\bibitem[Mai, 2025a]{mai2025bayesian}
	Mai, T.~T. (2025a).
	\newblock Bayesian pliable lasso with horseshoe prior for interaction effects
	in glms with missing responses.
	\newblock {\em arXiv preprint arXiv:2509.07501}.
	
	\bibitem[Mai, 2025b]{mai2025concentration}
	Mai, T.~T. (2025b).
	\newblock Concentration properties of fractional posterior in 1-bit matrix
	completion.
	\newblock {\em Machine Learning}, 114(1):7.
	
	\bibitem[Mai, 2025c]{mai2025handling}
	Mai, T.~T. (2025c).
	\newblock Handling bounded response in high dimensions: a horseshoe prior
	bayesian beta regression approach.
	\newblock {\em arXiv preprint arXiv:2505.22211}.
	
	\bibitem[Mai, 2025d]{mai2025hightobit}
	Mai, T.~T. (2025d).
	\newblock {High-dimensional Bayesian Tobit regression for censored response
		with Horseshoe prior}.
	\newblock {\em arXiv preprint arXiv:2505.08288}.
	
	\bibitem[Mai, 2025e]{mai2025properties}
	Mai, T.~T. (2025e).
	\newblock On properties of fractional posterior in generalized reduced-rank
	regression.
	\newblock {\em Journal of Multivariate Analysis}, 210:105481.
	
	\bibitem[Mai, 2025f]{mai2025sparse}
	Mai, T.~T. (2025f).
	\newblock A sparse pac-bayesian approach for high-dimensional quantile
	prediction.
	\newblock {\em Statistics and Computing}, 35(4):93.
	
	\bibitem[Mai, 2026]{mai2026optimal}
	Mai, T.~T. (2026).
	\newblock Optimal sparse phase retrieval via a quasi-bayesian approach.
	\newblock {\em Statistics and Computing}, 36(1):26.
	
	\bibitem[Parikh and Boyd, 2014]{parikh2014proximal}
	Parikh, N. and Boyd, S. (2014).
	\newblock Proximal algorithms.
	\newblock {\em Foundations and trends{\textregistered} in Optimization},
	1(3):127--239.
	
	\bibitem[Reinsel et~al., 2023]{reinsel2023multivariate}
	Reinsel, G.~C., Velu, R.~P., and Chen, K. (2023).
	\newblock {\em Multivariate Reduced-Rank Regression: Theory, Methods and
		Applications}, volume 225.
	\newblock Springer Nature.
	
	\bibitem[She and Chen, 2017]{she2017robust}
	She, Y. and Chen, K. (2017).
	\newblock Robust reduced-rank regression.
	\newblock {\em Biometrika}, 104(3):633--647.
	
	\bibitem[Tan et~al., 2024]{tan2024adaptive}
	Tan, X., Peng, L., Lian, H., and Liu, X. (2024).
	\newblock Adaptive huber trace regression with low-rank matrix parameter via
	nonconvex regularization.
	\newblock {\em Journal of Complexity}, 85:101871.
	
	\bibitem[Zhang, 2010]{zhang2010nearly}
	Zhang, C.-H. (2010).
	\newblock Nearly unbiased variable selection under minimax concave penalty.
	\newblock {\em The Annals of Statistics}, 38(2):894.
	
\end{thebibliography}
\end{document}